\documentclass[12pt]{article}
\usepackage{amssymb,latexsym}
\usepackage[dvips]{graphicx}
\newtheorem{theo}{Theorem}
\newtheorem{defi}[theo]{Definition}

\def\bea{\begin{eqnarray}}
\def\eea{\end{eqnarray}}
\def\nn{\nonumber}
\def\beq{\begin{equation}}
\def\eeq{\end{equation}}

\def\nn{\nonumber}

\def\C{\mathbb{C}}
\def\Z{\mathbb{Z}}
\def\R{\mathbb{R}}

\def\lb{[\![}
\def\rb{]\!]}

\renewcommand{\theequation}{\arabic{section}.\arabic{equation}}

\begin{document}
\begin{center}
{\Large \bf A classification of generalized quantum statistics}\\[4mm]
{\Large \bf associated with the exceptional Lie (super)algebras}\\[3cm]
{\bf N.I.\ Stoilova\footnote{Permanent address:
Institute for Nuclear Research and Nuclear Energy, Boul.\ Tsarigradsko Chaussee 72,
1784 Sofia, Bulgaria} and J.\ Van der Jeugt}\\[2mm]
Department of Applied Mathematics and Computer Science,\\
University of Ghent, Krijgslaan 281-S9, B-9000 Gent, Belgium.\\
E-mails: Neli.Stoilova@UGent.be, Joris.VanderJeugt@UGent.be.
\end{center}


\begin{abstract}
Generalized quantum statistics (GQS) associated to a Lie algebra or Lie superalgebra
extends the notion of para-Bose or para-Fermi statistics.
Such GQS have been classified for all classical simple Lie algebras and
basic classical Lie superalgebras.
In the current paper we finalize this classification for all exceptional
Lie algebras and superalgebras.
Since the definition of GQS is closely related to a certain $\Z$-grading
of the Lie (super)algebra $G$, our classification reproduces some known
$\Z$-gradings of exceptional Lie algebras. For exceptional Lie superalgebras
such a classification of $\Z$-gradings has not been given before.
\end{abstract}

\vspace{1cm}
\noindent
Running title: Classification for exceptional Lie (super)algebras\\[2mm]

\vspace{1cm}

\newpage
\renewcommand{\thesection}{\Roman{section}}
\renewcommand{\theequation}{\arabic{section}.{\arabic{equation}}}

\setcounter{equation}{0}
\section{Introduction} \label{sec:Introduction and classification method}%

More than 50 years ago, Green~\cite{Green} extended the notions of Bose and Fermi statistics 
to introduce so-called para-Bose and para-Fermi statistics. 
In para-statistics, the defining relations between operators which are (bilinear) 
commutators or anti-commutators for bosons and fermions are replaced by certain trilinear or triple relations.
Both for para-Bose and para-Fermi statistics, the defining relations can
be understood in terms of a Lie algebra or Lie superalgebra.
For example, the Lie algebra generated 
by $2n$ elements $F^\xi_i$ ($\xi=\pm$, $i=1,\ldots,n$) 
subject to the para-Fermi defining relations is $B_n=so(2n+1)$~\cite{KR}, 
and the Lie superalgebra generated by $2n$ odd elements 
$B^\xi_i$ ($\xi=\pm$, $i=1,\ldots,n$) subject to the para-Bose defining 
relations is the orthosymplectic Lie superalgebra $B(0|n)=osp(1|2n)$~\cite{Ganchev}. 

The examples of para-statistics~\cite{Green,Ohnuki} inspired others to
consider the notion of ``generalized quantum statistics'' (GQS) associated to
different classes of Lie algebras or superalgebras.
In particular, Palev~\cite{Palev1}-\cite{sl(1|n)} developed some of these extensions.
Using his examples and inspired by his definition of creation and annihilation 
operators in~\cite{Palev5}, a mathematical definition of ``generalized quantum
statistics'' was given in~\cite{GQS1}, followed by a complete classification of 
all classes of generalized quantum statistics for the classical 
Lie algebras $A_n$, $B_n$, $C_n$ and $D_n$~\cite{GQS1}.
In a second paper~\cite{GQS2}, this definition was given for the case of
Lie superalgebras, and a complete classification of all GQS for the basic 
classical Lie superalgebras $A(m|n)=sl(m+1|n+1)$,
$B(m|n)=osp(2m+1|2n)$, $C(n)=osp(2|2n-2)$ and $D(m|n)=osp(2m|2n)$ was obtained~\cite{GQS2}.
In the present paper, we complete this classification by giving all
GQS associated to the exceptional Lie algebras $G_2$, $F_4$, $E_6$, $E_7$, $E_8$ 
and to the exceptional Lie superalgebras $G(3)$, $F(4)$, $D(2,1; \alpha)$.

From the mathematical point of view, a GQS associated to a Lie (super)algebra $G$
is closely related to giving a certain $\Z$-grading of $G$ of the type 
$G= G_{-2} \oplus G_{-1} \oplus G_0 \oplus G_{+1} \oplus G_{+2}$. Such a 
$\Z$-grading is said to be of length~$5$ if $G_{\pm 2}\ne 0$; if $G_{\pm 2}=0$, but $G_{\pm1}\ne0$, then the
$\Z$-grading is of length~$3$.
These $\Z$-gradings also imply that one is dealing with Lie triple systems (in the 
case of Lie algebras, see~\cite{Jacobson}), or with Lie supertriple systems (in the
case of Lie superalgebras, see~\cite{Okubo}). Such triple systems have also
been referred to as ``(super)ternary algebras'', see~\cite{Bars}.
In this last paper, interesting examples and explicit constructions of ternary
algebras are given for many simple Lie algebras and Lie superalgebras. And in
section~VII of~\cite{Bars} one notes: ``It would be interesting to embark on a
complete classification of ternary algebras and superternary algebras and 
provide a list of all possible constructions of a given Lie (super)algebra from
(super)ternary algebras.'' As we shall see, the results of~\cite{GQS1} and~\cite{GQS2}
together with those of the present paper, provide such a complete classification.

We shall now recall the definition of a GQS, as given in~\cite{GQS1} and~\cite{GQS2},
and also briefly outline the classification method that will be followed.
This definition refers to the defining generators of $G$ as 
``creation and annihilation operators (CAOs) for $G$''.

Let $G$ be a Lie algebra or a basic Lie superalgebra, with bracket 
$\lb x,y \rb$, where (in $U(G)$) 
\[
\lb x,y \rb = x y - (-1)^{\deg(x)\deg(y)} y x,
\]
if $x$ and $y$ are homogeneous (for a Lie algebra, all elements are even and have degree~0;
for a Lie superalgebra, homogeneous elements are even or odd, having degree~0 or~1).
The following definition was given in~\cite{GQS1} and~\cite{GQS2}:
\begin{defi}
Let $G$ be a Lie algebra or a basic Lie superalgebra, with antilinear anti-involutive mapping $\omega$.
A set of $2N$ root vectors $x^\pm_i$ ($i=1,\ldots,N$) is called a set of
creation and annihilation operators for $G$ if:
\begin{itemize}
\item $\omega(x^\pm_i)=x^\mp_i$,
\item $G= G_{-2} \oplus G_{-1} \oplus G_0 \oplus G_{+1} \oplus G_{+2}$ is
a $\Z$-grading of $G$, with $G_{\pm 1}= \hbox{span}\{x^\pm_i,\ i=1\ldots,N\}$
and $G_{j+k}=\lb G_j,G_k \rb$.
\end{itemize}
The algebraic relations ${\cal R}$ satisfied by the operators $x_i^\pm$
are the relations of a generalized quantum statistics (GQS) associated with $G$.
\end{defi}

For a motivation of this definition, see~\cite{GQS1,GQS2}, where we also noted~\cite{GQS2} that 
this is a {\em mathematical} generalization of quantum statistics (in order to have
a quantum statistics in the physical sense, one should take into account 
additional requirements for the CAO's). 

A consequence of this definition is that $G$ is generated by $G_{-1}$ and $G_{+1}$,
i.e.\ by the set of CAOs, and since $G_{j+k}=\lb G_j,G_k \rb$, it follows that
\begin{equation}
G=\hbox{span}\{ x_i^\xi,\ \lb x_i^\xi,x_j^\eta \rb; \quad i,j=1,\ldots,N,\ \xi,\eta=\pm\}.
\end{equation} 
This implies that it is necessary and sufficient to give all relations of the following type:
\begin{itemize}
\item[(R1)] The set of all linear relations between the elements $\lb x_i^\xi, x_j^\eta\rb$ 
($\xi,\eta=\pm$, $i,j=1,\ldots,N$). 
\item[(R2)] The set of all triple relations of the form $\lb \lb x_i^\xi, x_j^\eta \rb,x_k^\zeta \rb=
\hbox{linear combination of }x_l^\theta$. 
\end{itemize}
So ${\cal R}$ consists of a set of quadratic relations and a set of triple relations.

The second condition in Definition~1 implies that $G_0$ itself is a subalgebra of $G$
spanned by root vectors of $G$. So $G_0$ is a regular subalgebra 
containing the Cartan subalgebra $H$ of $G$.
By the adjoint action, the remaining $G_i$'s are $G_0$-modules.
This has lead to the following technique in order to obtain a complete classification
of all GQS associated with $G$~\cite{GQS1,GQS2}:
\begin{enumerate}
\item
Determine all regular subalgebras $G_0$ of $G$. If not yet contained in $G_0$, replace $G_0$
by $G_0 + H$.
\item
For each regular subalgebra $G_0$, determine the decomposition of $G$
into simple $G_0$-modules $g_k$ ($k=1,2,\ldots$).
\item
Investigate whether there exists a $\Z$-grading of $G$ of the form
\begin{equation}
G=G_{-2} \oplus G_{-1} \oplus G_0 \oplus G_{+1} \oplus G_{+2}, 
\label{5grading}
\end{equation}
where each $G_i$ is either directly a module $g_k$ or else a sum of
such modules $g_1\oplus g_2\oplus \cdots$, such that
$\omega (G_{+i})=G_{-i}$.
\end{enumerate}

To find regular subalgebras one can use the method of deleting nodes from (extended) Dynkin 
diagrams~\cite{Dynkin,regular}. 
The second stage is straightforward by means of representation theoretical
techniques.
The third stage requires most of the work: one must try out all possible
combinations of the $G_0$-modules $g_k$, and see whether it is possible to
obtain a grading of the type~(\ref{5grading}). In this process, if one of the 
simple $G_0$-modules $g_k$ is such that $\omega(g_k)=g_k$, then it follows
that this module should be part of $G_0$. In other words, such a case
reduces essentially to another case with a larger regular subalgebra.

In the following sections we shall give a summary of the classification
process for exceptional Lie algebras and superalgebras.

\setcounter{equation}{0}
\section{The Lie algebras $G_2$ and $F_4$}

The Lie algebra $G_2$ of rank $2$ has dimension $14$. In terms of the orthonormal
vectors $\epsilon_1, \epsilon_2, \epsilon_3$ such that $\epsilon_1+\epsilon_2+\epsilon_3=0$,
the root system is given by
\beq
\Delta =\{ \epsilon_i - \epsilon_j, \;\epsilon_i+\epsilon_j-2\epsilon_k\;\;
(1\leq i\neq j\neq k\leq 3)\}.
\eeq 
The simple root system is
\beq
\Pi =\{\alpha_1= \epsilon_2 + \epsilon_3-2\epsilon_1, \;
\alpha_2=\epsilon_1-\epsilon_2\}
\eeq 
and the corresponding Dynkin diagram and extended Dynkin diagram are
given in Table~1. For $G_2$ and $F_4$ we follow the labeling and root systems
given in most of the textbooks, see e.g.~\cite{BMP,OV}.

The process described in the previous section, deleting nodes from the (extended)
Dynkin diagram, leads to the following results.

{\em Step 1}. Delete node~$1$ from the Dynkin diagram. The corresponding 
diagram is the Dynkin diagram of $G_0=sl(2)$. In this case, there 
are four $G_0$-modules and there exists a $\Z$-grading of length
$5$ with $G_{-1}$ and $G_{-2}$ chosen as follows (we give here the roots 
of the corresponding root vectors spanning $G_i$):
\beq
G_{-1} \rightarrow \{ \epsilon_1 - \epsilon_3, \;\epsilon_2 - \epsilon_3,\;
-\epsilon_2-\epsilon_3+2\epsilon_1, 
-\epsilon_1-\epsilon_3+2\epsilon_2\};
\eeq 
\beq
G_{-2} \rightarrow \{ 
\epsilon_1+\epsilon_2-2\epsilon_3 \}.
\eeq 

None of the other ways of deleting nodes from the Dynkin diagram or its extension
lead to any other results. So the classification leads to only one GQS, or in other
words to one gradings of the type~(\ref{5grading}) of length~5.

\vskip 5mm
Let us next consider the Lie algebra $F_4$, of rank $4$ and dimension $52$. 
In terms of the 
orthonormal vectors $\epsilon_1, \epsilon_2, \epsilon_3, \epsilon_4$ 
the root system is given by
\beq
\Delta =\{ \pm \epsilon_i \pm \epsilon_j\; (1\leq i\neq j\leq 4);\;\pm \epsilon_j\; (1\leq j\leq 4); \;
\frac{1}{2}(\pm \epsilon_1\pm \epsilon_2\pm \epsilon_3 \pm \epsilon_4) \}.
\eeq 
The simple root system is
\beq
\Pi =\{\alpha_1= \epsilon_2 - \epsilon_3, \;
\alpha_2=\epsilon_3-\epsilon_4, \; 
\alpha_3=\epsilon_4,\; 
\alpha_4=\frac{1}{2}(\epsilon_1- \epsilon_2- \epsilon_3 - \epsilon_4)    \}
\eeq 
and the corresponding Dynkin diagram and the extended Dynkin diagram are given in Table~1.
We consider now the various ways of deleting nodes from these diagrams, and investigate
whether they give rise to $\Z$-gradings of the type~(\ref{5grading}).

{\em Step 1}. Delete node $1$ from the Dynkin diagram. The corresponding 
diagram is the Dynkin diagram of $C_3=sp(6)$, so $G_0=H+C_3$. In this case, there 
are four $G_0$-modules and 
\[
F_4=G_{-2}\oplus G_{-1}\oplus G_0\oplus G_{+1}\oplus G_{+2}.
\] 
The corresponding roots of the root vectors belonging to $G_{-1}$ and
$G_{-2}$ can be chosen as follows:
\begin{eqnarray}
G_{-1} & \rightarrow &
\{ \epsilon_1, \epsilon_2, \epsilon_1 + \epsilon_3, \epsilon_1+\epsilon_4,
\epsilon_1 - \epsilon_3, \epsilon_1 - \epsilon_4, \epsilon_2 + \epsilon_3,
 \epsilon_2 + \epsilon_4, \epsilon_2 - \epsilon_3, \epsilon_2 - \epsilon_4,\nn\\
&& \frac{1}{2}(\epsilon_1+ \epsilon_2+ \epsilon_3 + \epsilon_4),
\frac{1}{2}(\epsilon_1+ \epsilon_2+ \epsilon_3 - \epsilon_4),
\frac{1}{2}(\epsilon_1+ \epsilon_2- \epsilon_3 + \epsilon_4),
\frac{1}{2}(\epsilon_1+ \epsilon_2- \epsilon_3 - \epsilon_4) \}, \nn\\
&&\nn\\
G_{-2} & \rightarrow &  \{ \epsilon_1+ \epsilon_2 \}. \nn
\end{eqnarray}

{\em Step 2}. Delete node $2$ or $3$ from the Dynkin diagram. The corresponding 
diagram is the Dynkin diagram of  $G_0=H+sl(2) \oplus sl(3)$. In this case, there 
are six $G_0$-modules but there exists no $\Z$-grading with the required 
properties. 

{\em Step 3}. Delete node $4$ from the Dynkin diagram. The corresponding 
diagram is the Dynkin diagram of $B_3=so(7)$, so $G_0=H+B_3$. In this case, there 
are four $G_0$-modules and 
\[
F_4=G_{-2}\oplus G_{-1}\oplus G_0\oplus G_{+1}\oplus G_{+2}.
\] 
$G_{-1}$ and $G_{-2}$ are determined by:
\begin{eqnarray}
G_{-1} & \rightarrow &
\{ \frac{1}{2}(\epsilon_1+ \epsilon_2+ \epsilon_3 + \epsilon_4),
\frac{1}{2}(\epsilon_1+ \epsilon_2+ \epsilon_3 - \epsilon_4),
\frac{1}{2}(\epsilon_1- \epsilon_2- \epsilon_3 + \epsilon_4),
\frac{1}{2}(\epsilon_1- \epsilon_2+ \epsilon_3 - \epsilon_4), \nn\\
&& 
\frac{1}{2}(\epsilon_1+ \epsilon_2- \epsilon_3 - \epsilon_4),
\frac{1}{2}(\epsilon_1- \epsilon_2+ \epsilon_3 + \epsilon_4),
\frac{1}{2}(\epsilon_1+ \epsilon_2- \epsilon_3 + \epsilon_4) \}, 
\frac{1}{2}(\epsilon_1- \epsilon_2- \epsilon_3 - \epsilon_4)\}, \nn\\
&&\nn\\ 
G_{-2} & \rightarrow & \{ \epsilon_1, \epsilon_1+ \epsilon_2,
\epsilon_1+ \epsilon_3, \epsilon_1+ \epsilon_4, \epsilon_1- \epsilon_2, 
\epsilon_1- \epsilon_3, \epsilon_1- \epsilon_4\}. \nn
\end{eqnarray}

{\em Step 4}. If we delete two or more nodes from the Dynkin diagram,
the resulting $\Z$-gradings of $F_4$ are no longer of the form
$F_4=G_{-2}\oplus G_{-1}\oplus G_0\oplus G_{+1}\oplus G_{+2}$. 

{\em Step 5}. Next, we move on to the extended Dynkin diagram of $F_4$. 
 If we delete one  node from the extended Dynkin diagram,
the resulting $\Z$-gradings of $F_4$ are no longer of the form
$F_4=G_{-2}\oplus G_{-1}\oplus G_0\oplus G_{+1}\oplus G_{+2}$. 

{\em Step 6}. Delete nodes $3$ and $4$ from the extended Dynkin diagram.
The remaining diagram is that of $\tilde{G}_0= sl(4)$. There are nine 
$\tilde{G}_0$-modules $g_k$, one of which is invariant under $\omega $
(say $g_1$). Then one must set $G_0=H+\tilde{G}_0 +g_1$, and one finds 
$G_0= H+B_3$. Now there are four $G_0$-modules and the considered case 
is isomorphic to that of step $3$.

In all other cases when we delete  two adjacent nodes 
from the extended Dynkin diagram there exists no $\Z$-grading with the required 
properties.   

{\em Step 7}. Delete nodes $1$ and $3$ from the extended Dynkin diagram.
The remaining diagram is that of $\tilde{G}_0= sl(2) \oplus sl(2) \oplus sl(2)$. 
There are nine 
$\tilde{G}_0$-modules $g_k$, one of which is invariant under $\omega $
(say $g_1$). Then one must set $G_0=H+\tilde{G}_0 +g_1$, and one finds 
$G_0= H+B_3$. Now there are four $G_0$-modules and again the considered case 
is isomorphic to that of step $3$.

{\em Step 8}. Delete nodes $1$ and $4$ from the extended Dynkin diagram.
The remaining diagram is that of $\tilde{G}_0= sl(2) \oplus B_2$. 
There are seven
$\tilde{G}_0$-modules $g_k$, one of which is invariant under $\omega $
(say $g_1$). Then one must set $G_0=H+\tilde{G}_0 +g_1$, and one finds 
$G_0= H+C_3$. Now there are four $G_0$-modules and  the considered case 
is isomorphic to that of step $1$.

In all other cases when we delete  two nonadjacent nodes 
from the extended Dynkin diagram there exists no $\Z$-grading with the required 
properties.  

{\em Step 9}. If we delete three or more nodes from the extended Dynkin 
diagram, the corresponding $\Z$-gradings of $F_4$ have no longer the required
properties.

In the case of $F_4$, we have explicitly spelled out all the steps of the
classification process, yielding finally only two different contributions
to our classification (from step~1 and step~3). This is essentially to show that
we have completed the whole classification work. In the following sections,
we shall no longer present the steps that reduce to one of the earlier found
contributions, but only present and describe the steps yielding new results.

\setcounter{equation}{0}
\section{The Lie algebra $E_6$}

For the remaining exceptional Lie algebras $E_6$, $E_7$ and $E_8$, our labeling
of simple roots is again the usual one. But our choice of (simple) roots in terms
of vectors $\epsilon_i$ is slightly different. For $E_8$, our choice is essentially
the same as in~\cite[Table~1]{OV}, except that we work with an independent
basis $\epsilon_i$ in $\R^8$ (and not a redundant basis in $\R^9$). The roots
for $E_8$ are the same as in~\cite{BMP} (but in this last textbook the choice of
simple roots is different). For $E_7$ and $E_6$ it is convenient for us to take
the same root space as for $E_8$. The simple roots of $E_7$ are then those of $E_8$
with the first one deleted, and the simple roots of $E_6$ are then those of $E_7$
with the first one deleted. 

The Lie algebra $E_6$ of rank $6$ has dimension $78$. We will use the following 
root system of $E_6$. Consider the $8$-dimensional real vector space $\R^8$
with orthonormal basis vectors $\epsilon_i \; (i=1,\ldots, 8)$. The roots of $E_6$ are elements
of the $6$-dimensional subspace $V$ of $\R^8$ consisting of those elements 
$\sum_{i=1}^8 c_i\epsilon_i$ with $c_1+c_2=0$ and $\sum_{i=3}^8 c_i=0$. A set of simple 
roots of $E_6$ is then given by the elements
\beq
\alpha_i = \epsilon_{i+2}-\epsilon_{i+3}\;\;(i=1,\ldots,5), \;\;\alpha_6=
\frac{1}{2}(\epsilon_1-\epsilon_2-\epsilon_3-\epsilon_4-\epsilon_5+\epsilon_6
+\epsilon_7+\epsilon_8).
\eeq
All $72$ nonzero roots are given by
\begin{eqnarray}
&& \pm(\epsilon_i-\epsilon_j), \;\; (1\leq i\neq j \leq 2 \hbox{ or } 3\leq i\neq j\leq 8)\nn \\
&& \frac{1}{2}(\sum_{i=1}^8(-1)^{a_i}\epsilon_i), \;\; (a_i\in \{ 0,1\};\;
\sum_{i=1}^2a_i=1,\;\; \sum_{i=3}^8 a_i=3).
\end{eqnarray}

The corresponding Dynkin diagram and the extended Dynkin diagram are given in Table~1.

{\em Step 1}. Delete node $1$ from the Dynkin diagram. The corresponding 
diagram is the Dynkin diagram of $D_5=so(10)$, so $G_0=H+D_5$. In this case, there 
are two $G_0$-modules and 
\[
E_6= G_{-1}\oplus G_0\oplus G_{+1}.
\] 
$G_{-1}$ itself is determined by:
\begin{eqnarray}
G_{-1} & \rightarrow  &
\{ \epsilon_1 - \epsilon_2;\; \epsilon_3 - \epsilon_i \;\; (4\leq i \leq 8); \nn\\
&& \frac{1}{2}(\epsilon_1- \epsilon_2+ \epsilon_3+ 
\sum_{i=4}^8 (-1)^{a_i}\epsilon_i)\;\; (a_i\in \{ 0,1\},\;
\sum_{i=4}^8 a_i=3) \}. \nn
\end{eqnarray}
It is easy to see that $\dim G_{-1}=16$.

\noindent
{\em Step 2}. Delete node $2$ from the Dynkin diagram. The corresponding 
diagram is the Dynkin diagram of $sl(2)\oplus sl(5)$, so $G_0=H+sl(2)\oplus sl(5)$. 
In this case, there 
are four $G_0$-modules and 
\[
E_6=G_{-2}\oplus G_{-1}\oplus G_0\oplus G_{+1}\oplus G_{+2}.
\] 
$G_{-1}$ and $G_{-2}$ are determined by:
\begin{eqnarray}
G_{-1} & \rightarrow &
\{ \epsilon_3 - \epsilon_i,\; \epsilon_4 - \epsilon_i \;\; (5\leq i \leq 8);\nn\\
&& \frac{1}{2}(\epsilon_1- \epsilon_2+ \epsilon_3 - \epsilon_4 + 
\sum_{i=5}^8 (-1)^{a_i}\epsilon_i)\;\; (a_i\in \{ 0,1\},\;
\sum_{i=5}^8 a_i=2); \nn\\
&& \frac{1}{2}(\epsilon_1- \epsilon_2- \epsilon_3 + \epsilon_4 + 
\sum_{i=5}^8 (-1)^{a_i}\epsilon_i)\;\; (a_i\in \{ 0,1\},\;
\sum_{i=5}^8 a_i=2) \}; \nn \\
G_{-2} & \rightarrow & \{ \epsilon_1- \epsilon_2 ;\nn\\
&& \frac{1}{2}(\epsilon_1- \epsilon_2+ \epsilon_3 + \epsilon_4 + 
\sum_{i=5}^8 (-1)^{a_i}\epsilon_i)\;\; (a_i\in \{ 0,1\},\;
\sum_{i=5}^8 a_i=3) \}. \nn
\end{eqnarray}
So $\dim G_{-1}=20$ and $\dim G_{-2}=5$.

\noindent
{\em Step 3}. Delete node $6$ from the Dynkin diagram. The corresponding 
diagram is the Dynkin diagram of $sl(6)$, so $G_0=H+sl(6)$. 
In this case, there 
are again four $G_0$-modules, 
\[
E_6=G_{-2}\oplus G_{-1}\oplus G_0\oplus G_{+1}\oplus G_{+2},
\] 
and $G_{-1}$ and $G_{-2}$ are determined by:
\begin{eqnarray}
G_{-1} & \rightarrow &
\{ \frac{1}{2}(\epsilon_1- \epsilon_2+ 
\sum_{i=3}^8 (-1)^{a_i}\epsilon_i)\;\; (a_i\in \{ 0,1\},\;
\sum_{i=3}^8 a_i=3) \}; \nn \\
G_{-2} &  \rightarrow &  \{ \epsilon_1- \epsilon_2\}. \nn
\end{eqnarray}
In this case $\dim G_{-1}=20$ and $\dim G_{-2}=1$.

\noindent
{\em Step 4}. Delete nodes $1$ and $5$ from the Dynkin diagram. The corresponding 
diagram is the Dynkin diagram of $D_4=so(8)$, so $G_0=H+so(8)$. 
In this case, there are six $G_0$-modules, but there is only one way in which these $G_0$-modules can be combined
so as to lead to a grading of the form
\[
E_6=G_{-2}\oplus G_{-1}\oplus G_0\oplus G_{+1}\oplus G_{+2}.
\] 
Now $G_{-1}$ and $G_{-2}$ are determined by:
\begin{eqnarray}
G_{-1} & \rightarrow &
\{ \epsilon_3- \epsilon_i,\; \epsilon_i- \epsilon_8\;\; (4\leq i\leq 7); \nn\\
&& \frac{1}{2}(\epsilon_1- \epsilon_2+ \epsilon_3+\epsilon_8 +
\sum_{i=4}^7 (-1)^{a_i}\epsilon_i)\;\; (a_i\in \{ 0,1\},\;
\sum_{i=4}^7 a_i=3); \nn\\
&& \frac{1}{2}(\epsilon_1- \epsilon_2- \epsilon_3-\epsilon_8 +
\sum_{i=4}^7 (-1)^{a_i}\epsilon_i)\;\; (a_i\in \{ 0,1\},\;
\sum_{i=4}^7 a_i=1) \}; \nn \\
G_{-2} & \rightarrow &
\{ \epsilon_1- \epsilon_2,\; \epsilon_3- \epsilon_8; \nn\\
&& \frac{1}{2}(\epsilon_1- \epsilon_2+ \epsilon_3-\epsilon_8 +
\sum_{i=4}^7 (-1)^{a_i}\epsilon_i)\;\; (a_i\in \{ 0,1\},\;
\sum_{i=4}^7 a_i=2) \}. \nn 
\end{eqnarray}
Now $\dim G_{-1}=16$ and $\dim G_{-2}=8$.

All other cases corresponding to the deletion of one, two or more
nodes from the Dynkin diagram give either cases isomorphic to Steps
$1$-$4$ or the corresponding $\Z$-gradings of $E_6$ have no longer the required 
properties. Also the analysis starting from the extended 
Dynkin diagram yields no new cases. So the above four cases complete
the classification for $E_6$.

\setcounter{equation}{0}
\section{The Lie algebra $E_7$}

\noindent
The roots of $E_7$ can also be described in the space $\R^8$. They are elements of the 
$7$-dimensional subspace $V'$ consisting of elements $\sum_{i=1}^8 c_i\epsilon_i $
with $\sum_{i=1}^8 c_i=0 $. A set of simple roots of $E_7$ consists of the six simple
roots $\alpha_i$ $(i=1,\ldots, 6)$ of $E_6$ plus the extra root 
\beq
\epsilon_2 -\epsilon_3.
\eeq 
The corresponding Dynkin diagram and the extended Dynkin diagram are given in Table~1.

By construction, the $E_6$ subsystem of $E_7$ is evident. The nonzero roots of 
$E_7$ consist of 
\begin{eqnarray}
&& \pm(\epsilon_i-\epsilon_j), \;\; (1\leq i < j \leq 8); \nn \\
&& \frac{1}{2}(\sum_{i=1}^8(-1)^{a_i}\epsilon_i), \;\; (a_i\in \{ 0,1\};\;
\sum_{i=1}^8 a_i=4) \label{rootsE7}.
\end{eqnarray}
Note that the 72 nonzero roots of $E_6$ are indeed part of the $126$ nonzero roots~(\ref{rootsE7}).

\noindent
{\em Step 1}. Delete node $1$ from the Dynkin diagram. The corresponding 
diagram is the Dynkin diagram of $E_6$, so $G_0=H+E_6$. In this case, there 
are two $G_0$-modules and 
\[
E_7= G_{-1}\oplus G_0\oplus G_{+1}.
\] 
The space $G_{-1}$ is determined by:
\begin{eqnarray}
G_{-1} & \rightarrow  &
\{ \epsilon_1 - \epsilon_i,\; \epsilon_2 - \epsilon_i \;\; (3\leq i \leq 8); \nn\\
&& \frac{1}{2}(\epsilon_1+ \epsilon_2+ \sum_{i=3}^8 (-1)^{a_i}\epsilon_i)\;\; (a_i\in \{ 0,1\},\;
\sum_{i=3}^8 a_i=4) \}. \nn
\end{eqnarray}
So $\dim G_{-1}=27$.

\noindent
{\em Step 2}. Delete node $2$ from the Dynkin diagram. The corresponding 
diagram is the Dynkin diagram of $sl(2)\oplus so(10)$, so $G_0=H+sl(2)\oplus so(10)$. 
In this case, there 
are four $G_0$-modules and 
\[
E_7=G_{-2}\oplus G_{-1}\oplus G_0\oplus G_{+1}\oplus G_{+2}.
\] 
Explicitly, the roots corresponding to $G_{-1}$ and
$G_{-2}$ can be chosen as follows:
\begin{eqnarray}
G_{-1} & \rightarrow &
\{ \epsilon_1 - \epsilon_2,\;  \epsilon_1-\epsilon_3;\; \nn\\
&& \epsilon_2 - \epsilon_i,\; \epsilon_3 - \epsilon_i \;\; (4\leq i \leq 8);\nn\\
&& \frac{1}{2}(\epsilon_1- \epsilon_2+ \epsilon_3 + 
\sum_{i=4}^8 (-1)^{a_i}\epsilon_i)\;\; (a_i\in \{ 0,1\},\;
\sum_{i=4}^8 a_i=3); \nn\\
&& \frac{1}{2}(\epsilon_1+ \epsilon_2- \epsilon_3 + 
\sum_{i=4}^8 (-1)^{a_i}\epsilon_i)\;\; (a_i\in \{ 0,1\},\;
\sum_{i=4}^8 a_i=3) \}; \nn \\
G_{-2} & \rightarrow & \{ \epsilon_1- \epsilon_i \;\;(4\leq i \leq 8);\nn\\
&& \frac{1}{2}(\epsilon_1+ \epsilon_2+ \epsilon_3 + 
\sum_{i=4}^8 (-1)^{a_i}\epsilon_i)\;\; (a_i\in \{ 0,1\},\;
\sum_{i=4}^8 a_i=4) \}. \nn
\end{eqnarray}
So $\dim G_{-1}=32$ and $\dim G_{-2}=10$.

\noindent
{\em Step 3}. Delete node $6$ from the Dynkin diagram. The corresponding 
diagram is the Dynkin diagram of $D_6=so(12)$, so $G_0=H+D_6$. 
Also in this case, there 
are four $G_0$-modules, $E_7$ has a $\Z$-grading of length~5, and $G_{-1}$ and
$G_{-2}$ are determined by:
\begin{eqnarray}
G_{-1} & \rightarrow &
\{ \epsilon_1 - \epsilon_i,\; \epsilon_i - \epsilon_8 \;\; (2\leq i \leq 7);\nn\\
&& \frac{1}{2}(\epsilon_1- \epsilon_8+ 
\sum_{i=2}^7 (-1)^{a_i}\epsilon_i)\;\; (a_i\in \{ 0,1\},\;
\sum_{i=2}^7 a_i=3) \}; \nn \\
G_{-2} &  \rightarrow &  \{ \epsilon_1- \epsilon_8\}. \nn
\end{eqnarray}
In this case $\dim G_{-1}=32$ and $\dim G_{-2}=1$.

\noindent
{\em Step 4}. Delete node $7$ from the Dynkin diagram. The corresponding 
diagram is the Dynkin diagram of $sl(7)$, so $G_0=H+sl(7)$. 
Again, there are four $G_0$-modules, $E_7$ has a $\Z$-grading of length~5, and
$G_{-1}$ and $G_{-2}$ are fixed by:
\begin{eqnarray}
G_{-1} & \rightarrow &
\{ \frac{1}{2}(\epsilon_1+ 
\sum_{i=2}^8 (-1)^{a_i}\epsilon_i)\;\; (a_i\in \{ 0,1\},\;
\sum_{i=2}^8 a_i=4) \}; \nn \\
G_{-2} &  \rightarrow &  \{ \epsilon_1 - \epsilon_i \;\; (2\leq i \leq 8)\}. \nn
\end{eqnarray}
So now $\dim G_{-1}=35$ and $\dim G_{-2}=7$.

All other cases corresponding to the deletion of nodes from the 
Dynkin diagram or its extension, reduce to one of the previous four
results. So these four cases are the only contribution to the
classification.

\setcounter{equation}{0}
\section{The Lie algebra $E_8$}

\noindent
In terms of the orthonormal vectors $\epsilon_i$, $i=1,\ldots,8$ the root system
of $E_8$ is given by 
\begin{eqnarray}
&& \pm \epsilon_i\pm \epsilon_j, \;\; (1\leq i < j \leq 8); \nn \\
&& \frac{1}{2}\sum_{i=1}^8 \xi_i\epsilon_i, \;\; \xi_i=\pm 1 \;
\hbox{and the number of } \xi_i=+1 \;
\hbox{is even}. \label{rootsE8}
\end{eqnarray} 
A set of simple roots of $E_8$ consists of the seven simple
roots $\alpha_i$ $(i=1,\ldots, 7)$ of $E_7$ plus the extra root 
\beq
-\epsilon_1 -\epsilon_2.
\eeq 
The corresponding Dynkin diagram and the extended Dynkin diagram are given in Table~1. 
 
\noindent
{\em Step 1}. Delete node $1$ from the Dynkin diagram. The corresponding 
diagram is the Dynkin diagram of $E_7$, so $G_0=H+E_7$. 
In this case, there 
are four $G_0$-modules and 
\[
E_8=G_{-2}\oplus G_{-1}\oplus G_0\oplus G_{+1}\oplus G_{+2}.
\] 
The roots corresponding to $G_{-1}$ and
$G_{-2}$ are: 
\begin{eqnarray}
G_{-1} & \rightarrow &
 \{ \epsilon_i+\epsilon_j \;\; (1\leq i<j\leq 8);\;\;
 \frac{1}{2}\sum_{i=1}^8(-1)^{a_i}\epsilon_i\;\; (a_i\in\{ 0,1\}, \; \sum_{i=1}^8a_i=2)\}\nn\\
G_{-2} &  \rightarrow &
 \{ 
 \frac{1}{2}\sum_{i=1}^8\epsilon_i \}. \nn
\end{eqnarray}  
So clearly, $\dim G_{-1}=56$ and $\dim G_{-2}=1$.
 
\noindent
{\em Step 2}. Delete node $7$ from the Dynkin diagram. The corresponding 
diagram is the Dynkin diagram of $D_7=so(14)$, so $G_0=H+so(14)$. 
Also now there are four $G_0$-modules, $E_8$ admits a $\Z$-grading
of length~5, and the roots corresponding to $G_{-1}$ 
are given by: 
\begin{eqnarray}
G_{-1} & \rightarrow &
\{ \epsilon_1 + \epsilon_8; \;  -\epsilon_1+\epsilon_i,\;  -\epsilon_i+\epsilon_8 \;\; (2\leq i \leq 7);\nn\\
&& \epsilon_k+\epsilon_l \;\; (2\leq k<l \leq 7);\nn\\
&& \frac{1}{2}(\epsilon_1+ \epsilon_8 + 
\sum_{i=2}^7 (-1)^{a_i}\epsilon_i)\;\; (a_i\in \{ 0,1\},\;
\sum_{i=2}^7 a_i=2); \nn\\
&& \frac{1}{2}(-\epsilon_1+ \epsilon_8 + 
\sum_{i=2}^7 (-1)^{a_i}\epsilon_i)\;\; (a_i\in \{ 0,1\},\;
\sum_{i=2}^7 a_i=3); \nn\\
&& \frac{1}{2}(-\epsilon_1- \epsilon_8 +
\sum_{i=2}^7 \epsilon_i) \}; \nn \\
G_{-2} & \rightarrow & \{ -\epsilon_1+\epsilon_8;\; \frac{1}{2}(\sum_{i=1}^{8}\epsilon_i); \;
\epsilon_i+\epsilon_8\;(2\leq i\leq 7); \nn\\
&&  \frac{1}{2}(-\epsilon_1 +\epsilon_8 +
\sum_{i=2}^7 (-1)^{a_i}\epsilon_i)\;\; (a_i\in \{ 0,1\},\;
\sum_{i=2}^7 a_i=1)\}. \nn
\end{eqnarray}
So $\dim G_{-1}=64$ and $\dim G_{-2}=14$.

All other cases corresponding to the deletion of nodes reduce to one of
the previous cases. So there are only two contributions in the
classification process.

The results of the classification for exceptional Lie algebras has been summarized
in Table~2. Note that the classification with $\Z$-gradings of length~5 ($\ell=5$
in Table~2) has been obtained before by means of different techniques, see~\cite{Kaneyuki}.
These results, including the cases with $\ell=3$ can also be found in~\cite[Chapter~II]{Faraut}.
Our classification presented here gives some extra information, namely the explicit
root structure of $G_{\pm 1}$ and $G_{\pm 2}$.

In the following sections our aim is to obtain a similar classification for
the exceptional Lie superalgebras.

\setcounter{equation}{0}
\section{The Lie superalgebra $D(2,1;\alpha )$}

For exceptional Lie superalgebras, we follow the notation and description of Kac~\cite{Kac}.
The description of the root system and a set of simple roots in terms of some basis
is that of~\cite{Kac,regular}. Note that for Lie superalgebras, not all sets of simple root systems 
are equivalent. So apart from the so-called ``distinguished'' set of simple roots
(and their corresponding Dynkin diagrams and their extensions), we should also
examine the nondistinguished sets of simple roots (with Dynkin diagrams and
their extensions). A good description of all such nondistinguished Dynkin diagrams
is given in~\cite{vdleur}.

The Lie superalgebras $D(2,1;\alpha )$ $(\alpha \in \C \backslash  \{ 0,-1 \})$ form a one-parameter
family of superalgebras of rank $3$ and dimension $17$. The even, respectively odd, roots of $D(2,1;\alpha )$ are
expressed in terms of linear functions $\epsilon_1, \epsilon_2, \epsilon_3$:
\begin{eqnarray}
&& \Delta_0=\{ \pm 2 \epsilon_i, \;\; i=1,2,3 \}, \nn \\
&& \Delta_1 = \{ \pm\epsilon_1\pm \epsilon_2\pm \epsilon_3\}  
\;\; {\rm \; (independent \;} \pm \;{\rm signs}). \label{rootsD}
\end{eqnarray} 
The distinguished set of simple roots is given by 
\beq
\{ \epsilon_1-\epsilon_2-\epsilon_3, \; 2\epsilon_2, \; 2\epsilon_3 \}.
\eeq 
The corresponding distinguished Dynkin diagram and the extended distinguished Dynkin diagram are
given in Table~3.
 
\noindent
{\em Step 1}. Delete node $1$ from the distinguished Dynkin diagram. The corresponding 
diagram is the Dynkin diagram of $sl(2)\oplus sl(2)$, so $G_0=H+ sl(2) \oplus sl(2)$. 
In this case, there 
are four $G_0$-modules and 
\[
D(2,1;\alpha )=G_{-2}\oplus G_{-1}\oplus G_0\oplus G_{+1}\oplus G_{+2}.
\] 
The corresponding roots of the root vectors belonging to $G_{-1}$ and
$G_{-2}$ can be chosen as follows: 
\begin{eqnarray}
&& G_{-1}  \rightarrow 
 \{ \epsilon_1+\epsilon_2+\epsilon_3, \; \epsilon_1-\epsilon_2+\epsilon_3,
 \; \epsilon_1+\epsilon_2-\epsilon_3,\; \epsilon_1-\epsilon_2-\epsilon_3 \}
 \nn\\
&&  G_{-2}  \rightarrow 
 \{ 
 2\epsilon_1 \}. \nn
\end{eqnarray}  
 
\noindent
{\em Step 2}. Delete node $2$ from the distinguished Dynkin diagram. The corresponding 
diagram is the Dynkin diagram of $sl(1|2)$, so $G_0=H+ sl(1|2) $. 
In this case, there 
are two $G_0$-modules and 
\[
D(2,1;\alpha )=G_{-1}\oplus G_0\oplus G_{+1}.
\] 
Now $G_{-1}$ is determined by: 
\begin{eqnarray}
&& G_{-1}  \rightarrow 
 \{ -2 \epsilon_1, \; -2 \epsilon_2,
 \; -\epsilon_1-\epsilon_2-\epsilon_3,\; -\epsilon_1-\epsilon_2+\epsilon_3 \}. \nn
\end{eqnarray}  
 
\noindent
{\em Step 3}. Delete nodes $2$ and $3$ from the distinguished Dynkin diagram. The corresponding 
diagram is the Dynkin diagram of $sl(1|1)$, so $G_0=H+ sl(1|1) $. 
There are six $G_0$-modules and three ways in which these $G_0$-modules can be combined. In all
these cases  
\[
D(2,1;\alpha )=G_{-2}\oplus G_{-1}\oplus G_0\oplus G_{+1}\oplus G_{+2}
\] 
with the roots corresponding to $G_{-1}$ and $G_{-2}$ as follows: 
\begin{eqnarray}
G_{-1} & \rightarrow &
 \{ 2 \epsilon_1, \; -2 \epsilon_2,
 \; \epsilon_1+\epsilon_2+\epsilon_3,\; -\epsilon_1-\epsilon_2+\epsilon_3 \}, \;
G_{-2} \rightarrow \{ \epsilon_1-\epsilon_2+\epsilon_3,\; 2\epsilon_3\};  \nn\\
G_{-1} & \rightarrow &
 \{ 2 \epsilon_1, \; -2 \epsilon_3,
 \; \epsilon_1+\epsilon_2+\epsilon_3,\; -\epsilon_1+\epsilon_2-\epsilon_3 \}, \;
G_{-2} \rightarrow \{ \epsilon_1+\epsilon_2-\epsilon_3,\; 2\epsilon_2\}; \nn\\
G_{-1} & \rightarrow &
 \{ 2 \epsilon_2, \; 2 \epsilon_3,
 \; \epsilon_1+\epsilon_2-\epsilon_3,\; \epsilon_1-\epsilon_2+\epsilon_3 \}, \;
G_{-2} \rightarrow \{ \epsilon_1+\epsilon_2+\epsilon_3,\; 2\epsilon_1\}. \nn 
\end{eqnarray}  

\noindent
{\em Step 4}.
All other cases corresponding to the deletion of 
nodes from the distinguished Dynkin diagram give either cases isomorphic to Steps~1-3 
or the corresponding $\Z$-gradings of $D(2,1;\alpha )$ have no longer the required 
properties. The investigation of the extended distinguished
Dynkin diagram gives the same result.
  
\noindent
{\em Step 5}. Next, one should repeat the process for all nondistinguished Dynkin
diagrams of $D(2,1;\alpha )$ and their extensions. This makes the work harder than the corresponding classification for the exceptional Lie algebras (which have only one Dynkin diagram and one extension).
We have repeated this procedure for all of them, leading to a lot of cases but not
leading to any new results (each case is isomorphic to one described already by means of the 
distinguished diagram).

\setcounter{equation}{0}
\section{The Lie superalgebra $G(3)$}

\noindent
The Lie superalgebra $G(3)$ of rank $3$ has dimension $31$.
The roots of $G(3)$ are given by
\begin{eqnarray}
&& \Delta_0=\{ \epsilon_j-\epsilon_k, \;\pm \epsilon_j; \; \pm 2\delta , \;j,k=1,2,3 \} \nn \\
&& \Delta_1 = \{ \pm(\epsilon_j+\delta),\; \pm(\epsilon_j- \delta),\; \pm\delta,   
\quad j=1,2,3 \}, \label{rootsG}
\end{eqnarray} 
where $\epsilon_j-\epsilon_k, \;\;\pm \epsilon_j$ are the roots of $G_2$ (satisfying
$\epsilon_1+\epsilon_2+\epsilon_3=0$) and $\pm2\delta$ are the roots of $sl(2)$ in $G(3)_{\bar 0}=
G_2\oplus sl(2)$. 
The distinguished set of simple roots is given by 
\beq
\{ \delta +\epsilon_1, \; \epsilon_2, \; \epsilon_3-\epsilon_2 \}.
\eeq 
The corresponding distinguished Dynkin diagram and the extended distinguished Dynkin diagram are given
in Table~3.

\noindent
{\em Step 1}. Delete node $1$ from the distinguished Dynkin diagram. The corresponding 
diagram is that of $G_2$, so $G_0=H+ G_2$. 
In this case, there 
are four $G_0$-modules and 
\[
G(3)=G_{-2}\oplus G_{-1}\oplus G_0\oplus G_{+1}\oplus G_{+2}.
\] 
The roots of $G_{-1}$ and $G_{-2}$ are given by: 
\begin{eqnarray}
&& G_{-1}  \rightarrow 
 \{ \delta, \; \epsilon_i+\delta, \; -\epsilon_i+\delta, 
 \; i=1,2,3 \}
 \nn\\
&&  G_{-2}  \rightarrow 
 \{ 
 2\delta \}. \nn
\end{eqnarray}  
 
 \noindent
{\em Step 2}. Delete node $3$ from the distinguished Dynkin diagram. The corresponding 
diagram is that of $sl(2|1)$, so $G_0=H+ sl(2|1)$. 
Also in this case, there are four $G_0$-modules, $G(3)$ has a $\Z$-grading of
length~5 and the roots of $G_{-1}$ and $G_{-2}$ are: 
\begin{eqnarray}
G_{-1} & \rightarrow &
 \{ \delta, \; -\epsilon_1, \epsilon_3, \; \delta+\epsilon_2, \; 
 \delta-\epsilon_2, \; -\epsilon_1+\epsilon_2, \; -\epsilon_2+\epsilon_3  \}
 \nn\\
G_{-2} & \rightarrow &
 \{ 
 2\delta, \delta-\epsilon_1, \; \delta+\epsilon_3,\; -\epsilon_1+\epsilon_3 \}. \nn
\end{eqnarray}  

\noindent
{\em Step 3}.
All other cases corresponding to the deletion of 
nodes from the distinguished Dynkin diagram give either cases isomorphic to Steps
1-2 or the corresponding $\Z$-gradings of $G(3)$ have no longer the required 
properties. The investigation of the extended distinguished
Dynkin diagram gives the same result.

\noindent
{\em Step 4}. Next, one should repeat the process for all nondistinguished Dynkin
diagrams of $G(3)$ and their extensions. 
The Dynkin diagram 
\[
\includegraphics{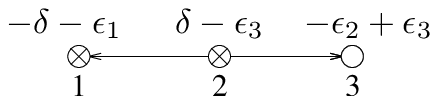}
\]
gives no new results, whereas its extension
\[
\includegraphics{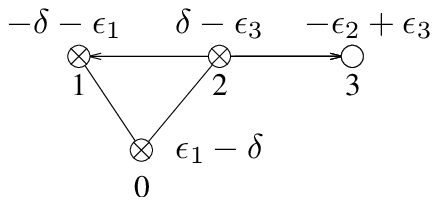}
\]
is interesting when one deletes node 1. Then the corresponding 
diagram is the Dynkin diagram of $sl(3|1)$, so $G_0=H+ sl(3|1) $. 
In this case, there are two $G_0$-modules and 
\[
G(3 )=G_{-1}\oplus G_0\oplus G_{+1}.
\] 
The roots of $G_{-1}$ are: 
\[
G_{-1}  \rightarrow 
 \{ \delta, \; \epsilon_1, \epsilon_2, \; \epsilon_3, \; 
 -2\delta, \; -\delta-\epsilon_1, \; -\delta-\epsilon_2, \;
 -\delta-\epsilon_3  \} .
\]

\noindent
{\em Step 5}. 
Also the Dynkin diagram 
\[
\includegraphics{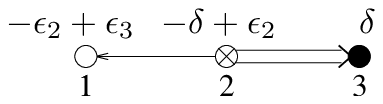}
\]
is interesting when one deletes node $1$. Then the corresponding 
diagram is that of $B(1|1)=osp(3|2)$, so $G_0=H+ B(1|1)$. 
In this case, there are four $G_0$-modules and $G(3)$ has a $\Z$-grading
of length~5. The roots of $G_{-1}$ and $G_{-2}$ are given by: 
\begin{eqnarray}
G_{-1} & \rightarrow &
 \{ \epsilon_1, \; -\epsilon_3, \epsilon_1-\delta, \; \epsilon_1+\delta, \; 
 -\epsilon_3-\delta, \; -\epsilon_3+\delta, \; \epsilon_1-\epsilon_2, \;
 \epsilon_2-\epsilon_3  \}
 \nn\\
G_{-2} & \rightarrow &
 \{ \epsilon_1-\epsilon_3 \}
\end{eqnarray}  

\noindent
{\em Step 6}. 
The extended Dynkin diagram 
\[
\includegraphics{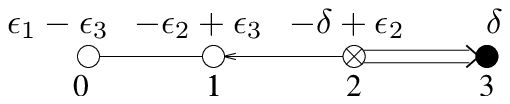}
\]
is interesting when one deletes node $2$. Then the corresponding 
diagram is the Dynkin diagram of $sl(3)\oplus B(0|1)=sl(3)\oplus osp(1|2)$, 
so $G_0=H+ sl(3)\oplus B(0|1) $. 
In this case, there are two $G_0$-modules and 
\[
G(3 )=G_{-1}\oplus G_0\oplus G_{+1}.
\] 
The roots belonging to $G_{-1}$ are given by: 
\[
G_{-1}  \rightarrow 
 \{ \epsilon_i, \; \epsilon_i+\delta, \; 
 \epsilon_i-\delta, \; i=1,2,3 \}.
\]  

\noindent
{\em Step 7}. The investigation of the other Dynkin diagrams and their extensions 
has not lead to any new results.

\setcounter{equation}{0}
\section{The Lie superalgebra $F(4)$}

\noindent
The Lie superalgebra $F(4)$ of rank $4$ has dimension $40$.
The roots of $F(4)$ are given by
\begin{eqnarray}
&& \Delta_0=\{ \pm \delta ; \pm\epsilon_i\; (1\leq i\leq 3); 
\;\pm \epsilon_i \pm \epsilon_j\; (1\leq i<j\leq 3) \} \nn \\
&& \Delta_1 = \{ \frac{1}{2}(\pm \delta \pm \epsilon_1\pm \epsilon_2 \pm \epsilon_3) \} 
 \label{rootsF}
\end{eqnarray} 
The distinguished set of simple roots is given by 
\beq
\{ \frac{1}{2}(-\epsilon_1 -\epsilon_2-\epsilon_3+\delta), \; \epsilon_3, \; 
\epsilon_2-\epsilon_3, \; \epsilon_1-\epsilon_2 \}.
\eeq 
The corresponding distinguished Dynkin diagram and the extended distinguished Dynkin diagram are
given in Table~3.

\noindent
{\em Step 1}. Delete node $1$ from the distinguished Dynkin diagram. The corresponding 
diagram is the Dynkin diagram of $B_3=so(7)$, so $G_0=H+ B_3 $. 
In this case, there are four $G_0$-modules and $F(4)$ has a $\Z$-grading of length~5. 
The roots of $G_{-1}$ and $G_{-2}$ are: 
\begin{eqnarray}
G_{-1}&  \rightarrow &
 \{ \frac{1}{2}(+\delta\pm\epsilon_1\pm\epsilon_2\pm\epsilon_3) \} \nn\\
G_{-2}&  \rightarrow &
 \{  \delta \}. \nn
\end{eqnarray}  
 
\noindent
{\em Step 2}. Delete node $3$ from the distinguished Dynkin diagram. The corresponding 
diagram is that of $sl(2|1)\oplus sl(2)$, so $G_0=H+ sl(2|1)\oplus sl(2) $. 
In this case, there are four $G_0$-modules and $F(4)$ has again a $\Z$-grading
of length~5.
The roots of $G_{-1}$ and $G_{-2}$ can be chosen as follows: 
\begin{eqnarray}
G_{-1} & \rightarrow &
 \{ \epsilon_i, \; \epsilon_i+\epsilon_3,\; \epsilon_i-\epsilon_3\; \; (i=1,2);\nn\\
 &&
 \frac{1}{2}(\epsilon_1-\epsilon_2-\epsilon_3+\delta), \; 
  \frac{1}{2}(\epsilon_1-\epsilon_2+\epsilon_3+\delta), \nn\\
 && -\frac{1}{2}(\epsilon_1-\epsilon_2+\epsilon_3-\delta), \;
 -\frac{1}{2}(\epsilon_1-\epsilon_2-\epsilon_3-\delta) \}, \; \nn\\
G_{-2} & \rightarrow &
 \{ \epsilon_1+\epsilon_2, \; \delta, \;  
  \frac{1}{2}(\epsilon_1+\epsilon_2-\epsilon_3+\delta),
  \frac{1}{2}(\epsilon_1+\epsilon_2+\epsilon_3+\delta)
  \}. \nn
\end{eqnarray}

\noindent
{\em Step 3}. Delete node $4$ from the distinguished Dynkin diagram. The corresponding 
diagram is that of the Lie superalgebra $C(3)=osp(2|4)$, so $G_0=H+C(3)$. 
In this case, there are two $G_0$-modules and 
\[
F(4)=G_{-1}\oplus G_0\oplus G_{+1}.
\] 
The roots of $G_{-1}$ can be chosen as follows: 
\begin{eqnarray}
G_{-1} & \rightarrow &
 \{ \epsilon_1, \; \delta, \; \epsilon_1+\epsilon_2,\; \epsilon_1-\epsilon_2, \;
 \epsilon_1+\epsilon_3, \; \epsilon_1-\epsilon_3,\nn\\
 &&
 \frac{1}{2}(\epsilon_1+\epsilon_2-\epsilon_3+\delta), \; 
  \frac{1}{2}(\epsilon_1+\epsilon_2+\epsilon_3+\delta), \nn\\
 && \frac{1}{2}(\epsilon_1-\epsilon_2-\epsilon_3+\delta), \;
 \frac{1}{2}(\epsilon_1-\epsilon_2+\epsilon_3+\delta) 
  \}. \nn
\end{eqnarray}

\noindent
{\em Step 4}. Delete nodes $1$ and $4$ from the extended distinguished Dynkin diagram. 
The corresponding 
diagram is the Dynkin diagram of $sl(2)\oplus B_2= sl(2) \oplus so(5)$ and $G_0=H+sl(2)\oplus B_2$. There are 
four $G_0$-modules and $F(4)$ has a $\Z$-grading of length~5, 
the roots of $G_{-1}$ and $G_{-2}$ being given by: 
\begin{eqnarray}
G_{-1} & \rightarrow &
 \{ 
 \frac{1}{2}(+ \epsilon_1\pm\epsilon_2\pm\epsilon_3\pm\delta)\}; \nn\\ 
G_{-2}& \rightarrow &
 \{ \epsilon_1,  \; \epsilon_1-\epsilon_2, \; \epsilon_1-\epsilon_3,
 \epsilon_1+\epsilon_2,\epsilon_1+\epsilon_3 \}. 
\end{eqnarray}

\noindent
{\em Step 5}. Delete nodes $3$ and $4$ from the extended distinguished Dynkin diagram. 
The corresponding 
diagram is that of $G_0=H+D(2,1;\alpha =-1/3)$.
Also in this case, there are four $G_0$-modules and $F(4)$ has a $\Z$-grading of length~5
with the roots of $G_{-1}$ and $G_{-2}$ given by: 
\begin{eqnarray}
G_{-1} & \rightarrow &
 \{ \epsilon_1, \; -\epsilon_2,  \; \epsilon_1\pm \epsilon_3,\; -\epsilon_2\pm\epsilon_3,\; 
 \frac{1}{2}(\epsilon_1-\epsilon_2\pm\epsilon_3\pm\delta)  \}; \nn\\
G_{-2} & \rightarrow &
 \{ \epsilon_1-\epsilon_2 \}. 
\end{eqnarray}

\noindent
{\em Step 6}.
All other cases corresponding to the deletion of
nodes from the distinguished Dynkin diagram or the extended distinguished 
Dynkin diagram give either cases isomorphic to Steps~1-5 
or the corresponding $\Z$-gradings of $F(4)$ have no longer the required 
properties. 

\noindent
{\em Step 7}. Next, one should repeat the process for all nondistinguished Dynkin
diagrams of $F(4)$ and their extensions. 
The Dynkin diagram 
\[
\includegraphics{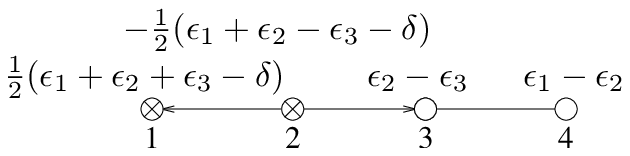}
\]
is interesting when one deletes node 1. Then the corresponding 
diagram is the Dynkin diagram of $sl(3|1)$, so $G_0=H+ sl(3|1) $. 
In this case, there also are four $G_0$-modules and $F(4)$
admits a $\Z$-grading of length~5. 
The roots of $G_{-1}$ and $G_{-2}$ are given by: 
\begin{eqnarray}
G_{-1} & \rightarrow &
 \{ \delta, \; \epsilon_1, \epsilon_2, \; \epsilon_3, \; 
 \frac{1}{2}(\epsilon_1-\epsilon_2+\epsilon_3+\delta), \; 
 \frac{1}{2}(\epsilon_1+\epsilon_2-\epsilon_3+\delta), \nn\\
 &&
 \frac{1}{2}(\epsilon_1+\epsilon_2+\epsilon_3-\delta), \;
 -\frac{1}{2}(\epsilon_1-\epsilon_2-\epsilon_3-\delta)
  \}
 \nn\\
G_{-2} & \rightarrow &
 \{ \epsilon_1+\epsilon_2, \; \epsilon_1+\epsilon_3, \;
 \epsilon_2+\epsilon_3,\; \frac{1}{2}(\epsilon_1+\epsilon_2+\epsilon_3+\delta)\}
\end{eqnarray}  

\noindent
{\em Step 8}. The investigation of the other Dynkin diagrams and their extensions 
has not lead to any new results.

The results of the classification for exceptional Lie superalgebras has been summarized
in Table~4. 
A number of interesting conclusions can be drawn from this table. 
First of all, note that all exceptional Lie superalgebras have a $\Z$-grading of
length~3 ($G(3)$ has even two such $\Z$-gradings). 
Secondly, most of these $\Z$-gradings of length~3 or~5 are not consistent with
the $\Z_2$-grading of the Lie superalgebra $G=G_{\bar 0}\oplus G_{\bar 1}$ itself
($G_{\bar 0}$ denoting the even elements and $G_{\bar 1}$ the odd elements). 
For $D(2,1;\alpha)$ and $G(3)$,
only the first grading given in Table~4 is a consistent grading. For $F(4)$ the 
first and the fourth grading in Table~4 are consistent, the others not. 

\setcounter{equation}{0}
\section{Example and conclusions}

Our analysis has led to a complete classification of all GQS associated with 
the exceptional Lie (super)algebras, and thus, together with~\cite{GQS1}
and~\cite{GQS2}, to a classification of all GQS associated with simple
Lie algebras and basic Lie superalgebras. The results have been
conveniently summarized in Tables~2 and~4.
In the terminology of~\cite{Bars}, we have given all possible ways of
constructing a simple Lie algebra or a basic Lie superalgebra from
a (super)ternary algebra, thus providing an answer to the question raised
in the introduction.

Since we have given only the root structure of the subspaces in the
grading~(\ref{5grading}), let us also give one example of the corresponding
GQS description in terms of the root vectors (or CAOs). 
For our example, consider $G=D(2,1;\alpha)$, with the grading of length~3 (second
line in Table~4). The root vectors corresponding to the roots $\{2\epsilon_1, 2\epsilon_2, 
\epsilon_1+\epsilon_2+\epsilon_3, \epsilon_1+\epsilon_2-\epsilon_3\}$ are denoted by,
respectively,
\begin{equation}
a_1^+,\ a_2^+,\ b_{+1}^+,\ b_{-1}^+.
\end{equation}
These are the ``creation operators'', $a_1^+$ and $a_2^+$ being even and $b_{+1}^+$ and
$b_{-1}^+$ being odd generators. These four operators span $G_{+1}$. The corresponding ``annihilation
operators'' spanning $G_{-1}$ are
\begin{equation}
a_1^-,\ a_2^-,\ b_{+1}^-,\ b_{-1}^-.
\end{equation}
Since the grading is of length~3, $\lb G_{+1},G_{+1}\rb=0$ and $\lb G_{-1},G_{-1}\rb=0$, so 
these elements mutually supercommute. 
The set ${\cal R}$ consists of all quadratic and triple relations. The quadratic relations are given by
\begin{eqnarray}
&& [a_i^+, a_j^-]=0, \hbox{ if } i\ne j; \nn \\
&& [a_1^{\pm},b_k^{\mp}] = [a_2^{\mp},b_k^{\pm}], \qquad (k=\pm 1); \label{D1} \\
&& 2\sigma_1 [a_1^-,a_1^+] + 2\sigma_2 [a_2^-,a_2^+] + \{b_{-1}^-,b_{+1}^+\} - \{b_{+1}^-,b_{-1}^+\}=0. \nn
\end{eqnarray}
Herein,
\[
(\sigma_1,\sigma_2,\sigma_3)=(1+\alpha,-1,-\alpha)
\]
are labels that are often used to describe the algebras $D(2,1;\alpha)$~\cite{D21a}.
The triple relations read:
\begin{eqnarray}
&& [ [ a_i^+,a_i^-],a_j^\pm] = \pm 2 \delta_{ij} a_j^\pm ;\qquad (i,j\in\{1,2\}) \nn \\
&& [ [ a_i^+,a_i^-],b_k^\pm] = \pm b_k^\pm ; \qquad (i=1,2; k=\pm 1) \nn \\
&& [ \{ b_i^+,b_j^-\},a_k^\pm] = \mp (j-i)\sigma_k a_k^\pm ; \qquad (i,j\in\{-1,+1\}; k\in\{1,2\}) \label{D2} \\
&& [ \{ b_i^+,b_j^-\},b_k^\xi] = -\,\alpha\,\delta_{\xi,+}(k-i)b^+_{i+j+k} -
\alpha\,\delta_{\xi,-}(k-j)b^-_{i+j+k}; \nn \\
&& \qquad\qquad\qquad (i,j,k\in\{-1,+1\}; \xi=\pm). \nn
\end{eqnarray}

Note that a set of CAOs together with a complete set
of relations ${\cal R}$ unambiguously describes the Lie superalgebra. 
So~(\ref{D1})-(\ref{D2}) gives a complete description of $D(2,1;\alpha)$
in terms of 8 generators subject to quadratic and triple relations.
In fact, each case of our classification gives the description of an exceptional Lie (super)algebra
in terms of a number of generators subject to certain relations.
This can also be reformulated in terms of the notion of Lie (super)triple 
systems~\cite{Jacobson,Okubo}. Following the definition, the subspace $G_{-1}\oplus G_{+1}$
(i.e.\ the subspace spanned by all CAOs) is a Lie (super)triple system for
the universal enveloping algebra $U(G)$.

As an application for the results on exceptional Lie superalgebras, 
we mention the possible solutions of Wigner Quantum Systems~\cite{WQS1}.
Roughly speaking, the compatibility conditions to be satisfied by a Wigner Quantum Oscillator
system (see formula (3.7) in~\cite{WQS2}) are written in terms of certain odd operators $A_i^\pm$; furthermore,
these compatibility conditions are special triple relations~\cite{WQS3}. 
For a possible candidate solution all the CAOs of ${\cal R}$ should be odd
operators. In other words, we should select those GQSs from our classification with
a consistent $\Z$-grading. Using Table~4, this is easy exercise, 
that has already been discussed at the end of the previous section.

\bigskip
\noindent{\bf Acknowledgments}
\medskip

\noindent

N.I.\ Stoilova was supported by a project from the Fund for Scientific Research -- Flanders (Belgium).

\newpage
\noindent
{\bf Table 1}.
Exceptional Lie algebras, their (extended) Dynkin diagrams with a labeling of
the nodes and the corresponding simple roots.
\vskip 1mm
\noindent
\begin{tabular}{lc}
\hline
LA & Dynkin diagram and extended Dynkin diagram \\
\hline \\
$G_2$ & \includegraphics{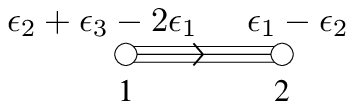}\qquad  \includegraphics{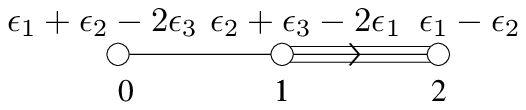}\\ \hline \\
$F_4$ & \includegraphics{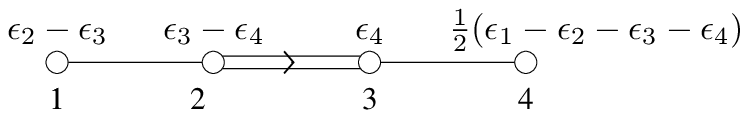}\\
      & \includegraphics{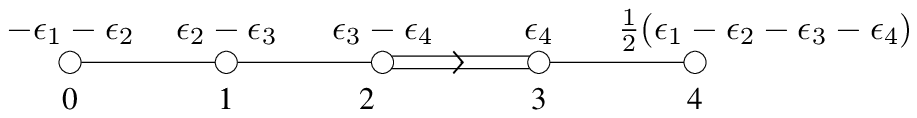}\\ \hline \\
$E_6$ & \includegraphics{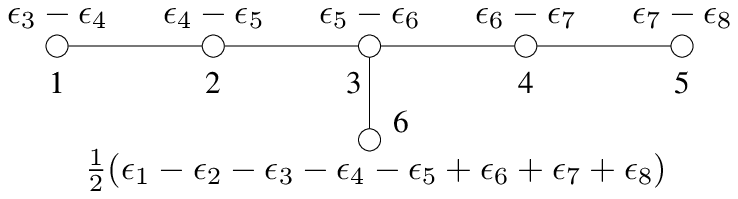}\\
      & \includegraphics{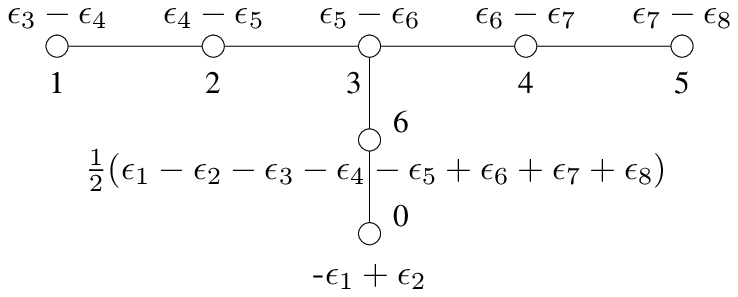}\\ \hline \\
$E_7$ & \includegraphics{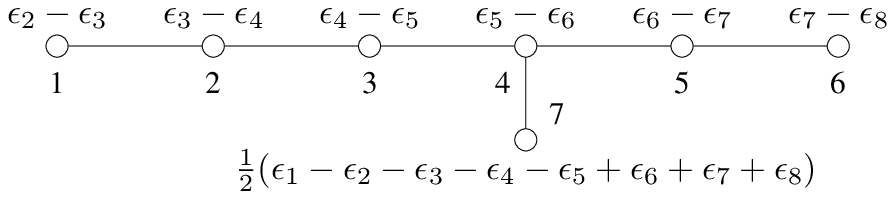}\\
      & \includegraphics{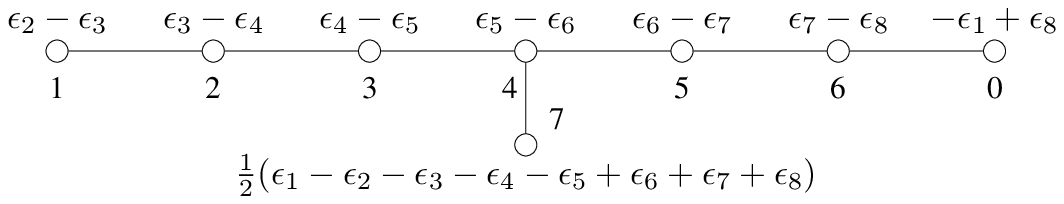}\\ \hline \\
$E_8$ & \includegraphics{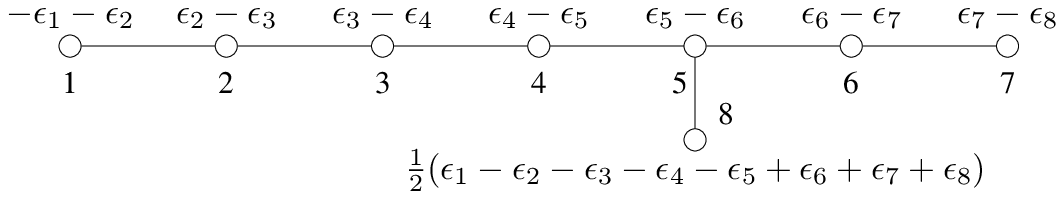}\\
      & \includegraphics{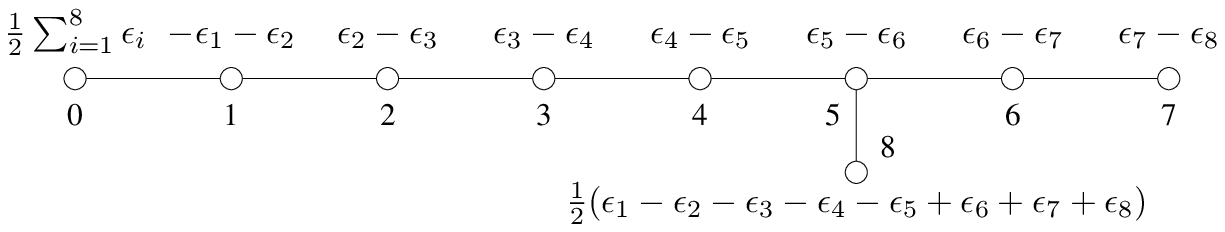}\\
\hline
\end{tabular}

\newpage
\noindent
{\bf Table 2}.
Summary of the classification for exceptional Lie algebras: all nonisomorphic GQS 
(or all the different $\Z$-gradings of type~(\ref{5grading})) are given. 
For each case, we list: the subalgebra $G_0$; 
the length $\ell$ of the $\Z$-grading (3 or 5); 
$\dim G_0$; $\dim G_{-1}=\dim G_{+1}$, which is also the number $N$ of creation or annihilation operators;
and $\dim G_{-2}=\dim G_{+2}$.
\vskip 2mm
\noindent
\begin{tabular}{l|l|r|r|r|r}
\hline
LA          & $G_0$ & $\ell$ & $\dim G_0$ & $\dim G_{1}$ & $\dim G_2$   \\
\hline \hline
$G_2$  &  $\C \oplus sl(2)$ &  5 & 4 & 4 & 1\\
\hline  
$F_4$  & $\C \oplus sp(6)$ &  5 & 22 & 14 & 1\\
       & $\C \oplus so(7)$ &  5 & 22 & 8 & 7\\
\hline  
$E_6$  & $\C \oplus so(10)$ &  3 & 46 & 16 & 0\\
       & $\C \oplus sl(2)\oplus sl(5)$ &  5 & 28 & 20 & 5\\
       & $\C \oplus sl(6)$ &  5 & 36 & 20 & 1\\
       & $\C \oplus \C \oplus so(8)$ &  5 & 30 & 16 & 8\\
\hline       
$E_7$  & $\C \oplus E_6$ &  3 & 79 & 27 & 0\\
       & $\C \oplus sl(2)\oplus so(10)$ &  5 & 49 & 32 & 10\\
       & $\C \oplus so(12)$ &  5 & 67 & 32 & 1\\
       & $\C \oplus sl(7)$ &  5 & 49 & 35 & 7\\
\hline       
$E_8$  & $\C \oplus E_7$ &  5 & 134 & 56 & 1\\
       & $\C \oplus so(14)$ &  5 & 92 & 64 & 14\\
\hline\hline         
\end{tabular}

\newpage
\noindent
{\bf Table 3}.
Exceptional Lie superalgebras, their distinguished (extended) Dynkin diagrams with a labeling of
the nodes and the corresponding simple roots.
\vskip 1mm
\noindent
\begin{tabular}{lc}
\hline
LSA & distinguished Dynkin diagram and extended distinguished Dynkin diagram \\
\hline \\
$D(2,1;\alpha)$ & \includegraphics{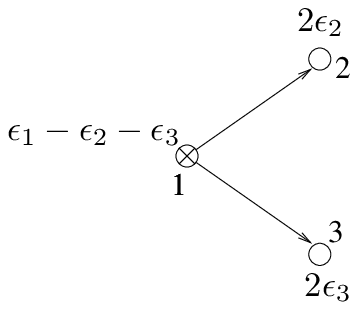}\qquad \includegraphics{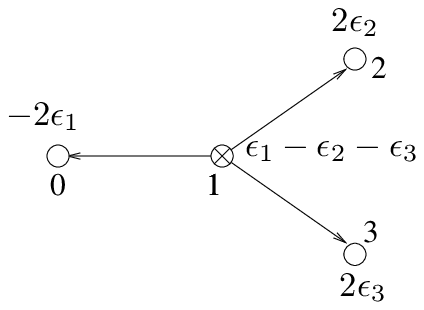}\\ \hline \\
$G(3)$ & \includegraphics{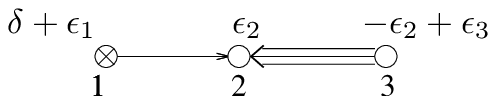}\qquad \includegraphics{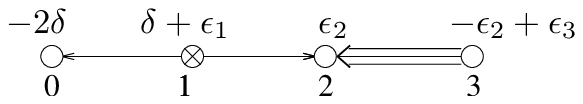}\\ \hline \\
$F(4)$ & \includegraphics{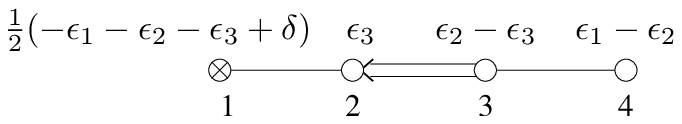}\\
      & \includegraphics{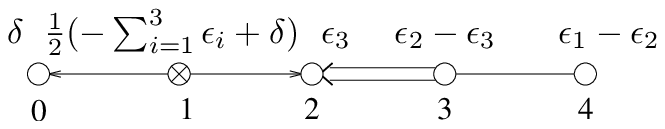}\\ 
\hline
\end{tabular}

\newpage
\noindent
{\bf Table 4}.
Summary of the classification for exceptional Lie superalgebras: all nonisomorphic GQS 
(or all the different $\Z$-gradings of type~(\ref{5grading})) are given. 
For each case, we list: the subalgebra $G_0$; 
the length $\ell$ of the $\Z$-grading (3 or 5); 
$\dim G_0$; $\dim G_{-1}=\dim G_{+1}$, which is also the number $N$ of creation or annihilation operators;
and $\dim G_{-2}=\dim G_{+2}$. We write these dimensions in the form $a+b$, where $a$ is the dimension of the even
part of $G_i$ and $b$ is the dimension of the odd part of $G_i$.
\vskip 2mm
\noindent
\begin{tabular}{l|l|r|r|r|r}
\hline
LSA          & $G_0$ & $\ell$ & $\dim G_0$ & $\dim G_{1}$ & $\dim G_2$   \\
\hline \hline
$D(2,1;\alpha)$  &  $\C \oplus sl(2)\oplus sl(2)$ &  5 & 7+0 & 0+4 & 1+0\\
       &  $\C\oplus sl(1|2)$ &  3 & 5+4 & 2+2 & 0+0\\
       &  $\C\oplus\C\oplus sl(1|1)$ &  5 & 3+2 & 2+2 & 1+1\\
\hline  
$G(3)$  & $\C \oplus G_2$ &  5 & 15+0 & 0+7 & 1+0\\
       & $\C \oplus sl(1|2)$ &  5 & 5+4 & 4+3 & 2+2\\
       & $sl(3|1)$ &  3 & 9+6 & 4+4 & 0+0\\
       & $\C \oplus osp(3|2)$ &  5 & 7+6 & 4+4 & 1+0\\
       & $sl(3)\oplus osp(1|2)$ &  3 & 11+2 & 3+6 & 0+0\\
\hline  
$F(4)$  & $\C \oplus so(7)$ &  5 & 22+0 & 0+8 & 1+0\\
       & $\C \oplus sl(1|2)\oplus sl(2)$ &  5 & 8+4 & 6+4 & 2+2\\
       & $\C\oplus osp(2|4)$ &  3 & 12+8 & 6+4 & 0+0\\
       & $\C \oplus sl(2)\oplus so(5)$ &  5 & 14+0 & 0+8 & 5+0\\
       & $\C\oplus D(2,1;-1/3)$ &  5 & 10+8 & 6+4 & 1+0\\
       & $\C \oplus sl(3|1)$ &  5 & 10+6 & 4+4 & 3+1\\
\hline\hline         
\end{tabular}

\end{document}